\documentclass[twocolumn,floatfix]{aastex631}
\usepackage[latin9]{inputenc}
\usepackage{graphicx}
\usepackage{hyperref}
\newcommand{\fnl}{f_{\rm NL}}

\begin{document}
\title{Impact of Large-Scale Anisotropies on Galaxy Clustering and Cosmological Constraints}
\author[0000-0001-7888-4270]{Prabhakar Tiwari}
\affil{Department of Physics, Guangdong Technion - Israel Institute of Technology, Shantou, Guangdong 515063, P.R. China}
\begin{abstract}
We critically assess the impact of significant dipole and large-scale anisotropies on galaxy clustering signals, with a focus on radio continuum surveys. Our study reveals that these anisotropies---resulting from intrinsic cosmological effects and/or observational systematics---profoundly influence the two-point correlation function (2PCF) and angular power spectrum ($C_\ell$). Notably, large-scale anisotropies can obscure or simulate non-Gaussianity signals, complicating the extraction of precise cosmological information. 
The results emphasize that it is crucial to address systematics and rigorously mask the dipole and its surrounding multipoles to obtain accurate cosmological constraints. This approach is essential for extracting cosmological results from clustering signals, particularly for future surveys such as SKA, DESI, and LSST, to ensure the precision and reliability of cosmological analyses. 
\end{abstract}
\section{Introduction}
\label{sc:intro}
Galaxy clustering is a key tool for probing cosmological parameters, offering critical insights into the matter distribution and evolution of the Universe \citep{Peebles:1980}. By comparing observed clustering patterns with theoretical models, we impose constraints on fundamental parameters such as matter density, fluctuation amplitude, and structure growth. This process refines our understanding of the Universe's composition and the mechanisms behind galaxy and large-scale structure formation. In the era of precision cosmology, where achieving exceptional accuracy in parameter measurements is crucial, we rely on advanced observational techniques and data analysis methods. Next-generation surveys, such as the Square Kilometre Array (SKA; \citealt{Wilkinson:1991,Bacon:2020}), the Dark Energy Spectroscopic Instrument (DESI; \citealt{DESI:2016,DESI:2022}), and the Large Synoptic Survey Telescope (LSST; \citealt{Ivezic:2019LSST}), are set to dramatically enhance our ability to map the Universe and explore phenomena like dark energy, modified gravity \citep{Clifton:2012}, and primordial non-Gaussianities \citep{Komatsu:2001}. These advancements will critically evaluate the current cosmological model and explore alternative cosmological models.

However, these analyses face challenges due to large-scale anisotropies in galaxy distributions. Whether arising from intrinsic cosmological effects or observational systematics, these anisotropies can distort clustering signals and complicate accurate cosmological inference. A prime example is the observed dipole in radio and infrared galaxy counts, first identified in the NRAO VLA Sky Survey (NVSS) \citep{Blake:2002, Singal:2011}. This dipole has generated debate over its origin-whether it reflects our motion relative to the cosmic rest frame, an intrinsic large-scale structure feature, or uncorrected systematics \citep{Gibelyou:2012, Rubart:2013, Tiwari:2014ni,Tiwari:2016adi, Colin:2017,Dolfi:2019, Tiwari:2019TGSS, Secrest:2022}. The impact of such anisotropies is profound. The dipole and other large-scale modes can obscure the isotropic Gaussian assumption commonly used in standard analyses, complicating the detection of subtle effects such as primordial non-Gaussianity or scale-dependent biases from inflationary models \citep{Matarrese:2000,Dalal:2008,Bruni:2012}. 

Most leading cosmological clustering studies have thus far overlooked these effects, operating under the assumption of large-scale isotropy and homogeneity---commonly known as the Cosmological Principle, a fundamental basis for standard cosmological analyses. Yet recent observations, especially from radio and infrared surveys \citep{Tiwari:2014ni,Tiwari:2016adi, Colin:2017, Secrest:2020CPQ, Secrest:2022,Singal:2024,vonHausegger:2024}, reveal a prominent dipole signal in galaxy distributions, making it crucial to consider this signal and underscoring the urgency of addressing this issue.

In this letter, we investigate the effect of large-scale anisotropies on galaxy clustering measurements, focusing on the two-point correlation function (2PCF) and the angular power spectrum ($C_\ell$). Using NVSS data as a case study, we assess implications for upcoming surveys like SKA, DESI, and LSST. We highlight the necessity of aggressive masking of large-scale anisotropies and rigorous systematic treatment to mitigate the dipole and adjacent multipoles, which can otherwise introduce significant biases in cosmological parameter estimates. We demonstrate how anisotropies can mimic or obscure key cosmological signals, such as non-Gaussianities and scale-dependent biases. We need for robust data processing techniques and careful management of large-scale systematics to fully exploit the capabilities of next-generation surveys. We conclude that effectively addressing large-scale anisotropies is essential for achieving precise cosmological inferences.

\section{Galaxies as Cosmological Tracers}
\label{sc:tracers}

A typical galaxy catalog used to explore cosmology consists of the two-dimensional or three-dimensional positions of galaxies with specific properties, mapped over a fraction of the sky and across a range of radial distances---referred to as redshifts in cosmology. These catalogs are carefully prepared to ensure completeness above a certain flux or luminosity threshold, resulting in what is known as a flux-limited catalog. Alternatively, one can select a subsample over a specific range of redshifts where the catalog is considered complete in terms of containing all galaxies within that range. These are known as volume-limited catalogs, which are somewhat challenging to achieve with our current observational capabilities.

Regardless of the galaxy sample used, covering a particular range of dark matter halo masses is crucial, as it ensures that the derived galaxy bias for the population in the catalog is well-defined. This is important because galaxies serve as biased tracers of the underlying matter distribution \citep{Kaiser:1984, Cole:1989,Dekel:1987,Peacock:2000}. To relate the background matter distribution---essentially the cosmological structure---using galaxy positions, a halo-galaxy biasing model is required \citep{Press:1974,Sheth:1999}. This model connects the observed distribution of galaxies to that of dark matter halos, which are linked to the underlying matter density field. By understanding and applying the appropriate biasing model, one can infer the dark matter distribution from the observed galaxy distribution, thereby gaining essential insights into the large-scale structure of the Universe and advancing our understanding of cosmology.

A real galaxy survey covers only a fraction of the sky.  Combining different catalogs can extend the sky area covered \citep{Colin:2017, Tiwari:2019l123}; however, the galactic plane remains inaccessible due to high dust extinction, stellar crowding, and galactic contamination. Additionally, there are other masked regions where data is either contaminated by bright extended sources or where the survey's RMS noise is high. At best, large-scale clustering signals are typically studied with 70-90\% sky coverage. Nonetheless, the point remains that  methods must be employed to recover the full-sky clustering signal from surveys available only over a partial sky region. This recovery is particularly challenging if the dipole or any other multipole is significantly larger in comparison to its neighboring multipoles. In such cases, the power from a multipole with high amplitude leaks significantly into its neighboring multipoles, leading to notably higher clustering in those neighboring multipoles as well.

\section{Clustering Signal in the Presence of a Large Dipole} 
\label{sc:clustering}
Radio continuum surveys lack redshift information, so clustering analysis is typically performed using only the angular positions of galaxies. Conventional approaches involve calculating the two-point correlation function and the angular power spectrum to constrain cosmological parameters. These basic tests can be supplemented with higher-order correlations to achieve a more detailed understanding of the clustering signal and the underlying cosmology. 

It is well-established that galaxy catalogs from radio continuum and infrared surveys, and likely from visible surveys as well, exhibit a significant dipole \citep{Secrest:2022,Aluri:2022, Peebles:2022}. This dipole introduces complexities that can affect the reliability of clustering analyses. An excess dipole, whether arising from intrinsic cosmological effects or observational systematics, can contaminate clustering measurements and complicate the extraction of accurate cosmological information. In addition to a significant dipole, non-Gaussian initial perturbations can significantly contribute to large-scale clustering. A simple model for non-Gaussianity can be followed from \citep{Komatsu:2001}, where a local type of non-Gaussianity is described by the ``simplest weak nonlinear coupling" case, given as:
\begin{equation}
\label{eq:phi_fnl}
\Phi = \phi_g - \fnl(\phi_g^2 - \langle \phi_g^2 \rangle)
\end{equation}
where $\Phi$ represents the  primordial curvature perturbations, $\phi_g$ is the Gaussian perturbation, and $\fnl$ is the non-Gaussianity parameter. This leads to a scale-dependent galaxy bias \citep{Matarrese:2000,Dalal:2008,Seljak:2009,Desjacques:2010,Xia:2010NVSS,Bruni:2012,Maartens:2013}:
\begin{equation}
\label{eq:fnl_bias}
b(k,z) = b_g(z) + \fnl [b_g(z) - 1] \frac{3 \delta_{\text{ec}} \Omega_m H_0^2}{c^2 k^2 T(k) D(z)},
\end{equation}
where $b(k, z)$ is the scale- and redshift-dependent bias, with $k$ representing the wave number that corresponds to the inverse of the spatial scale of perturbations, and $z$ is the redshift, which denotes the epoch of evaluation. Here, $b_g(z)$ is the scale-independent Gaussian bias, $\delta_{\text{ec}}$ is the critical matter overdensity for ellipsoidal collapse, taken as $\delta_{\text{ec}} = 1.68 \sqrt{0.75} \approx 1.45$ \citep{Xia:2010NVSS, Maartens:2013}, $T(k)$ is the transfer function, $D(z)$ is the growth factor, $c$ is the speed of light, and $H_0$ is the Hubble constant. In this letter, while we have specifically considered the local-type primordial non-Gaussianity model for simplicity, it is to note that other inflationary-motivated non-Gaussianity models can similarly lead to scale-dependent biases in galaxy clustering \citep{Matarrese:2008non-Gaussianity}. 

The most widely used method to study the clustering signal from radio continuum surveys is conventionally through the 2-point correlation functions and the angular power spectrum ($C_\ell$). The theoretical formulation for the angular power spectrum is given by \citep{Peebles:1980}:
\begin{equation}
\label{eq:clth}
C_\ell = \frac{2}{\pi} \int dk \, k^2 \, P(k) \left\vert \int_{0}^{\infty} dz \, D(z) \, b(z, k) \, N(z) \, j_\ell(kr) \right\vert^2 \;,
\end{equation}
where $P(k)$ is the matter power spectrum, $N(z) \, dz$ is the radial distribution function, representing the probability of observing a galaxy between redshift $z$ and $(z + dz)$, and $j_\ell(kr)$ is the spherical Bessel function of the first kind for integer $\ell$. The galaxy bias $b(z, k)$ becomes scale-dependent in the presence of non-Gaussianity, i.e., when $\fnl \neq 0$. For Gaussian primordial perturbations, the galaxy bias is typically assumed to be scale-independent. Subsequently, the theoretical estimate of the 2-PCF, $w(\theta)$, derived from the angular power spectrum $C_\ell$:
\begin{equation}
\label{eq:cl2wth}
w(\theta) = \frac{1}{4 \pi} \sum_\ell (2\ell + 1) C_\ell \, \mathcal{P}_\ell(\cos \theta) \;,
\end{equation}
where $\mathcal{P}_\ell(x)$ represents the Legendre polynomials.

When recovering the full-sky clustering signal from surveys that cover only a partial sky, the dipole signal leaks into other nearby multipoles. This leakage can overlap with the non-Gaussianity signal, which is also dominant at large scales, making it challenging to distinguish between the two effects. To illustrate this issue, we present a typical case for the NRAO VLA Sky Survey (NVSS; \citealt{Condon:1998}).  We assume the $N(z)$ and bias $b_g(z)$ values for NVSS from \cite{Adi:2015nb}, i.e. $N(z) \propto z^{0.74} \exp\left[ - \left(\frac{z}{0.71} \right)^{1.1} \right]$  and $b_g(z) = 0.33 z^2+ 0.85z +1.6$. Here, we focus on qualitative plots; similar results are anticipated for other surveys, such as the upcoming SKA, DESI, and LSST. 

\begin{figure}
    \centering
    \includegraphics[width=\linewidth]{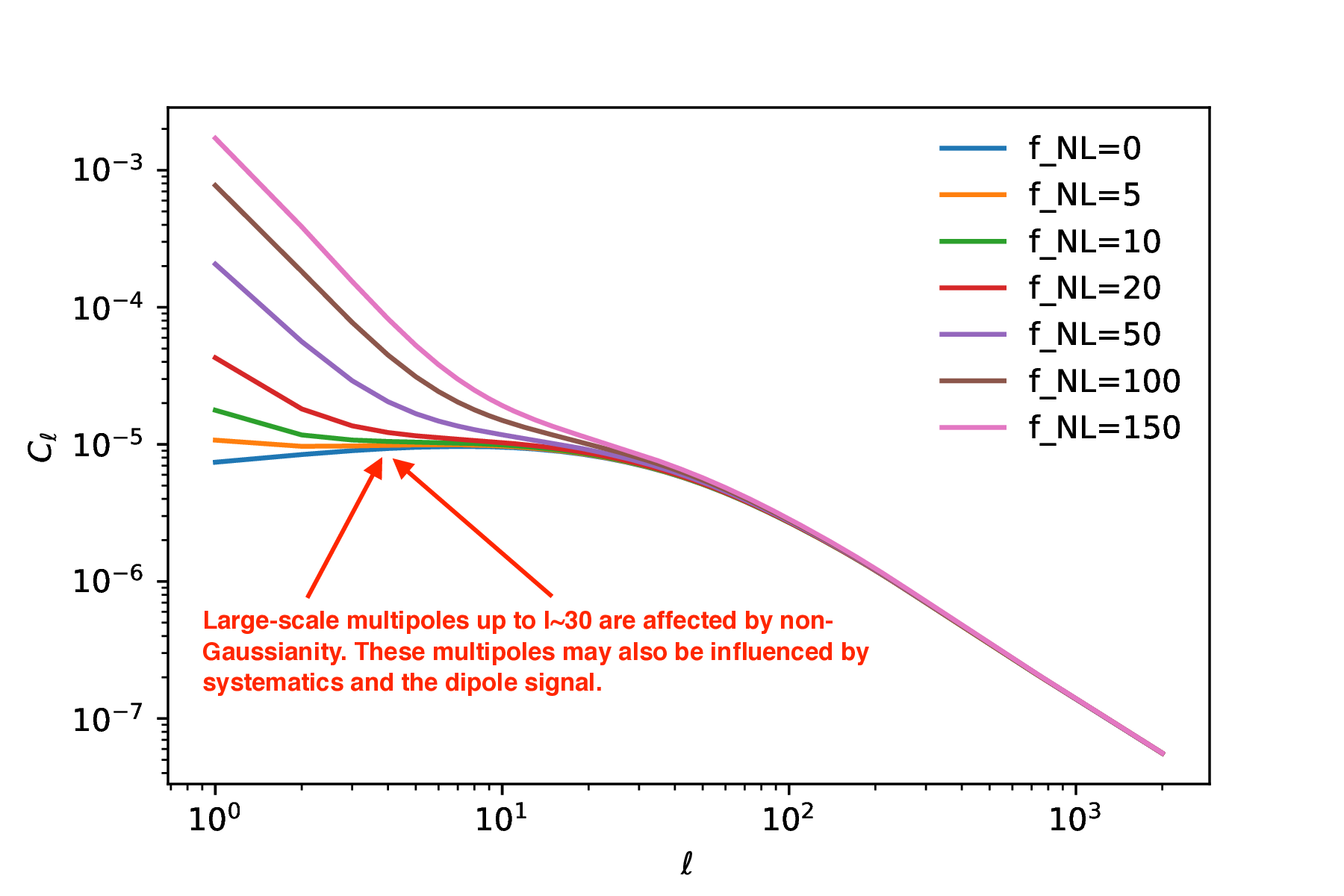}
    \caption{NVSS angular power spectrum as calculated using Equation \ref{eq:clth}. The effects of non-Gaussianity are incorporated by considering the scale-dependent bias as given in Equation \ref{eq:fnl_bias}. The $\fnl =0$ case represents the Gaussian perturbation scenario.}
    \label{fig:Cls}
\end{figure}

In Figure \ref{fig:Cls}, we present the angular power spectrum, which is widely used to study the clustering signal from radio continuum surveys. We consider the $\fnl$ parameter up to 150, which is the maximum expected value for this parameter \citep{Becker:2012}, although the current limit from CMB temperature and E-mode polarization maps is  $\fnl = -0.9 \pm 5.1$. For NVSS, \cite{Xia:2010NVSS} analyzed the clustering signal and found $\fnl = 62 \pm 27$. From the figure, it is evident that the non-Gaussianity signals can cause a significant increase in $C_\ell$s up to $\ell \approx 30$. However, at this scale, systematic effects may also produce similar increases in clustering. For instance, the TIFR GMRT Sky Survey (TGSS) ADR1 catalog exhibits increased clustering up to $\ell \approx 30$, likely due to systematics \citep{Tiwari:2019TGSS}. If these systematics are not properly accounted for, they may be misinterpreted as significant non-Gaussianity. Additionally, a prominent dipole signal-whether from systematic effects or cosmic origins-can further complicate the analysis. When recovering $C_\ell$s from partial sky coverage, which is always the case, the dipole signal can leak into nearby multipoles and mimic non-Gaussianity signals.
\begin{figure}
    \centering
    \includegraphics[width=\linewidth]{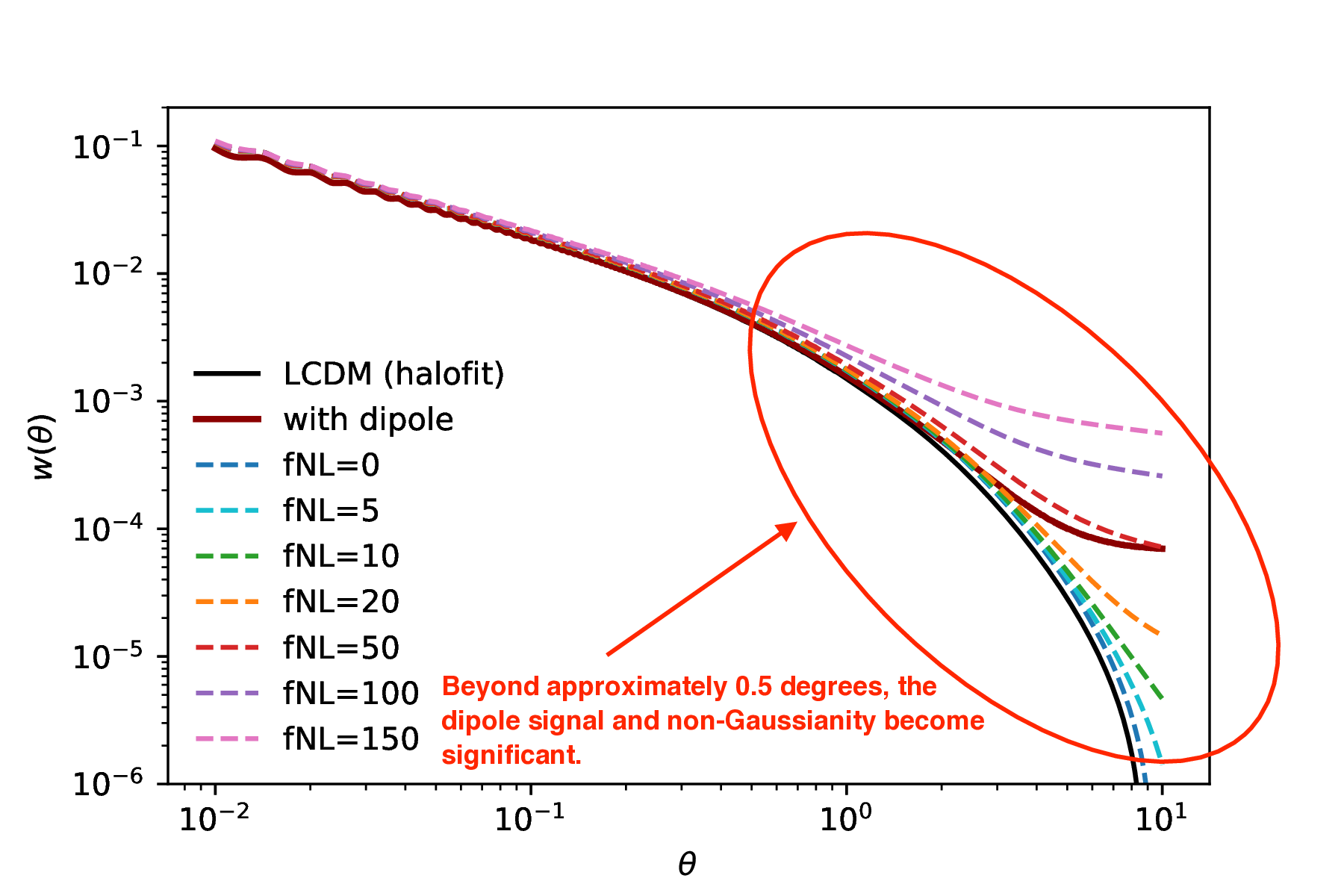}
    \caption{Estimates of the angular 2PCF in the presence of non-Gaussianity (non-zero $\fnl$) or dipole anisotropy ($|D| = 1.5 \times 10^{-2}$).}
    \label{fig:2PCF}
\end{figure}

To more clearly demonstrate how both the dipole signal and non-Gaussianity can produce similar clustering results, we present the angular 2PCF results in Figure \ref{fig:2PCF}. The 2PCF beyond scales of approximately 0.5 degrees is significantly influenced by non-Gaussianity. Interestingly, the presence of a dipole signal closely resembles the non-Gaussianity signal. Notably, the observed NVSS dipole, about $1.5 \times 10^{-2}$ \citep{Singal:2011, Gibelyou:2012, Rubart:2013, Tiwari:2014ni, Tiwari:2016adi, Dolfi:2019, Siewert:2020CRD, Secrest:2022}, corresponds approximately to an $\fnl$ value of 50, which is consistent with the observed $\fnl$ value within one sigma as reported by \cite{Xia:2010NVSS}. It appears that the dipole signal present in the data is being misinterpreted as a non-Gaussianity signal. Furthermore, since large-scale anisotropy and/or the presence of a non-Gaussianity signal affects the 2PCF over a significant range of angular scales, 2PCF is less favored for fitting cosmological models and obtaining cosmological results, particularly when dealing with non-Gaussianity signals. Before fitting any cosmological model to the 2PCF, the dipole signal should be removed. The simplest way to achieve this could be by fitting a cosine function with a constant to the observed 2PCF \footnote{Note from equation \ref{eq:cl2wth} that the dipole term corresponds to $\ell=1$ and is proportional to the cosine.}. The $C_\ell$s could be a better choice; however, it is essential to ensure that the data is not affected by large-scale systematics, and the dipole and its neighboring multipoles are masked before fitting cosmology. The results we presented are specific to NVSS, but as other radio continuum surveys have similar redshift profiles and galaxy biasing, we expect similar results. 

For spectroscopic and photometric surveys where redshift information is available, cosmology is studied using three-dimensional correlation functions, power spectrum, and bispectrum. However, as the dipole signal and non-Gaussianity signals both affect large scales, they remain ambiguous to distinguish even with these estimators. There are also general relativistic (GR) corrections that become larger at higher redshifts and are again more significant at large scales \citep{Bruni:2012}. However, since the galaxy population from most existing and planned surveys peaks below or around redshift 1, and considering that $\fnl$ is small as suggested by Planck observations, the GR and non-Gaussianity effects remain at the sub-percentage level. The dipole, other large-scale multipoles, and systematics remain the main challenges and must not be misinterpreted as non-Gaussianity signals. These large-scale signals need to be carefully removed before drawing any cosmological conclusions. 

We conclude that extracting cosmology from galaxy surveys in the presence of large-scale anisotropies is non-trivial and requires careful consideration. Firstly, systematics should be thoroughly studied and accounted for, as they can lead to an increase in the clustering signal at large scales \citep{Tiwari:2019TGSS}. Systematic effects present in the data may closely resemble non-Gaussianity signals in both the 2PCF and $C_\ell$s and three-dimensional power spectrum and other estimators. Not only non-Gaussianity, but even a dipole signal---whether due to systematics or a true cosmological origin---is difficult to distinguish when using 2PCF as a clustering measure. It may be somewhat distinguishable with $C_\ell$s if the survey data covers a sufficiently large sky area and the dipole signal does not leak significantly into nearby multipoles. General relativistic effects, non-Gaussianity, and possibly the cosmic dipole also exhibit some dependence on redshift \citep{Bruni:2012,Tiwari:2022SP}. Studying large-scale clustering signals in redshift slices may potentially help. However, even in this case, the signals could still be misinterpreted as mild non-Gaussianity or vice versa. It is advisable to use the CMB-extracted values of non-Gaussianity as a strong prior when interpreting large-scale anisotropy signals. Furthermore, when possible, the remaining cosmological parameters can be reliably extracted from galaxy catalogs by analyzing the clustering signal while avoiding the largest scales.

\bibliography{master}
\end{document}